\documentclass{webofc}
\usepackage[varg]{txfonts}   
\begin{document}
\title{The GRAND project and GRANDProto300 experiment}
%
%

\author{\firstname{Olivier} \lastname{Martineau-Huynh}\inst{1,2}\fnsep\thanks{\email{omartino@in2p3.fr}} for the GRAND collaboration
}

\institute{Sorbonne Universit\'e, Universit\'e Paris Diderot, Sorbonne Paris Cit\'e, CNRS,\\
Laboratoire de Physique Nucl\'eaire et de Hautes Energies (LPNHE), 4 place Jussieu, F-75252, Paris Cedex 5, France \\
\and
National Astronomical Observatories, Chinese Academy of Sciences, Beijing 100012, China \\
}

\abstract{
  The Giant Array for Neutrino Detection (GRAND) is a proposal for a giant observatory of ultra-high energy cosmic particles (neutrinos, cosmic rays and gamma rays). It will be composed of twenty subarrays of 10\,000 antennas each, totaling a detection area of 200\,000\,km$^2$. GRAND will reach unprecedented sensitivity to neutrinos allowing to detect cosmogenic neutrinos while its sub-degree angular resolution will also make it possible to hunt for point sources and possibly start neutrino astronomy. Combined with its gigantic exposure to ultra-high energy cosmic rays and gamma rays, GRAND will be a powerful tool to solve the century-long mistery of the nature and origin of the particles with highest energy in the Universe. On the path to GRAND,  the GRANDProto300 experiment will be deployed in 2020 over a total area of 200\,km$^2$. It primarly aims at validating the detection concept of GRAND, but also proposes a rich science program centered on a precise and complete measurement of the air showers initiated by cosmic rays with energies between 10$^{16.5}$ and 10$^{18}$\,eV, a range where we expect to observe the transition between the Galactic and extra-galactic origin of cosmic rays.
}
\maketitle
\section{Introduction}
\label{intro}
The Giant Radio Array for Neutrino Detection (GRAND) will be a network of 20 subarrays of $\sim$10\,000 radio antennas each, deployed in mountainous and radio-quiet sites around the world, totaling a combined area of 200\,000\,km$^2$. It will form an observatory of unprecedented sensitivity for ultra-high energy cosmic particles (neutrinos, cosmic rays and gamma rays). Here we first detail the GRAND detection concept, its science case and experimental challenges. In a second part we detail the GRANDProto300 experiment, a pathfinder for GRAND, but also an appealing scientific project on its own.

\section{The GRAND project}
\label{GRAND}
\subsection{Detection concept}
\label{concept}
\begin{figure}[h]
\centering
\includegraphics[width=10cm,clip]{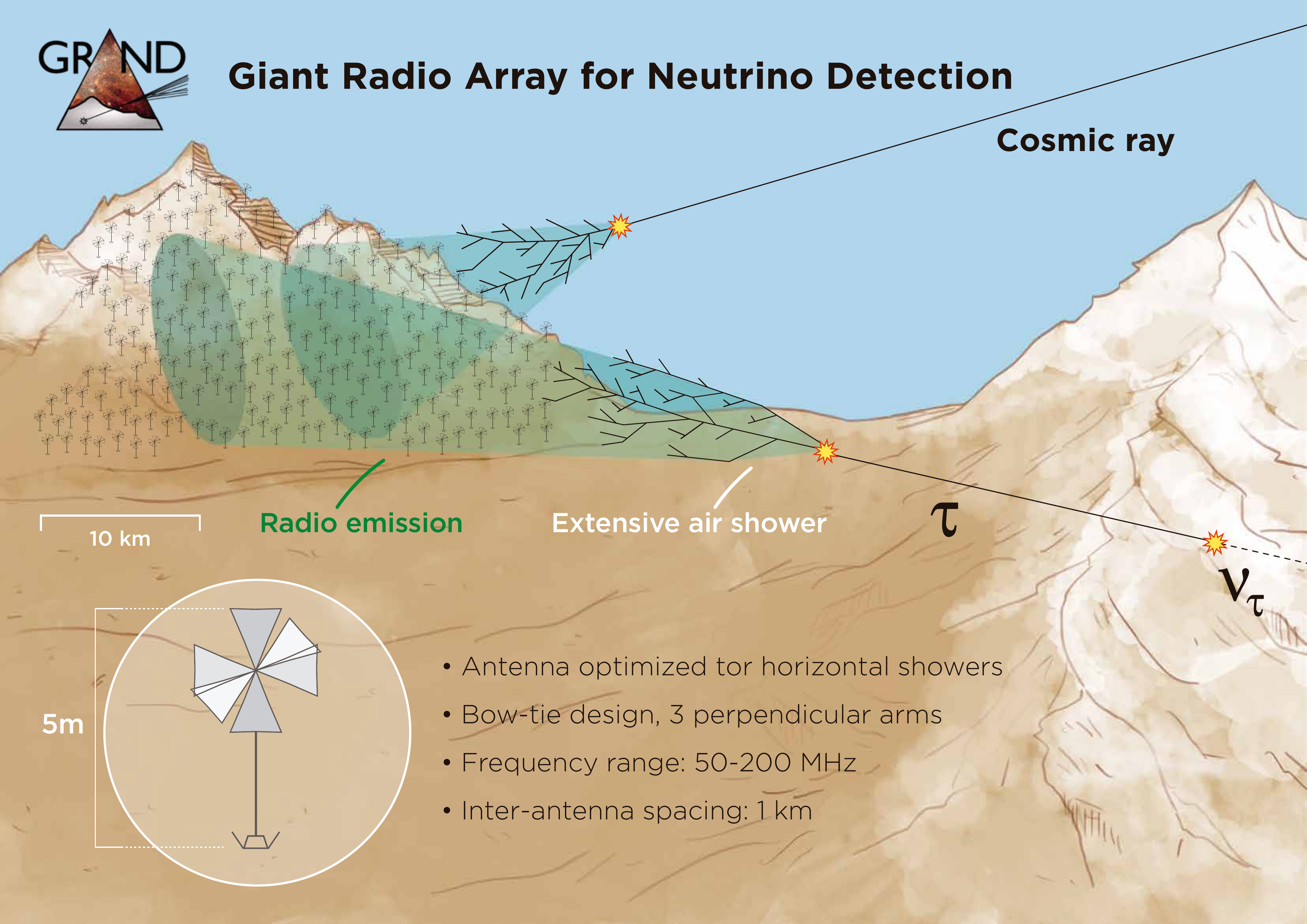}
\caption{GRAND detection principle for cosmic rays or gammas (detection of the EAS induced by the direct interaction of the cosmic particles in the atmosphere) and neutrinos (underground interaction with subsequent decay of the tau lepton in the atmosphere)}
\label{principle}
\end{figure}

Principles of radio detection of air showers are detailed in \cite{Huege:2016veh,Schroder:2016hrv}, and the GRAND detection concept is presented in \cite{WP}. It is briefly summarized below, and also illustrated in figure \ref{principle}. 

When it enters the Earth atmosphere, a cosmic particle may interact with air particles to induce an extensive air shower (EAS), which in turn, generates electromagnetic radiations mainly through the deflection by the Earth magnetic field of the charged particles composing the shower\,\cite{Kahn206}. This so-called {\it geomagnetic effect} is coherent in the tens of MHz frequency range, generating short (<1\,$\mu$s), transient electromagnetic pulses, with amplitudes large enough to allow for the detection of the EAS\,\cite{Allan:1971,Ardouin:2005qe,Falcke:2005tc} if the primary particle's energy is above 10$^{17}$\,eV typically. 

Cosmic neutrinos however have a very small probability of being detected through this process because of their tiny interaction cross-section with air particles. Yet, a tau neutrino can produce a tau lepton under the Earth surface through charged-current interactions. Thanks to its large range in rock and short lifetime, it may emerge in the Earth atmosphere and eventually decay to induce a detectable EAS\,\cite{Fargion:2000iz}. The Earth opacity to neutrinos of energies above 10$^{17}$\,eV however implies that only Earth-skimming trajectories allow for such a scenario.

This peculiarity, which can first be seen as a handicap for detection, turns out to be an asset for radiodetection: because of relativistic effects, the radio emission is indeed strongly beamed forward in a cone which opening is given by the Cerenkov angle $\theta_C\sim1^{\circ}$. For air shower trajectories close to the zenith, this induces a radio footprint at ground of few hundred meters diameter, requiring a large density of antennas at ground for a good sampling of the signal. For very inclined trajectories however, the larger distance of the antennas to the emission zone and the projection effect of the signal on ground combine to generate a much larger footprint\,\cite{Huege:2016veh}. Targeting air showers with very inclined trajectories ---either up-going for Earth-skimming neutrinos, or down-going for cosmic rays and gammas--- make it possible to detect them with a sparse array (typically one antenna per km$^2$). This is a key feature of the GRAND detector.  

Another driver in GRAND is to aim at mountainous areas with favorable topographies as deployment sites. An ideal topography consists of two opposing mountain ranges, separated by a few tens of kilometers. One range acts as a target for neutrino interactions, while the other acts as a screen on which the ensuing radio signal is projected. Simulations (see section \ref{simu}) show that such configurations result in a detection efficiency improved by a factor $\sim$4 compared to a flat site.

\subsection{Detector performances}
\label{simu}
\subsubsection{Neutrino sensitivity}

\begin{figure}[h]
\centering
\includegraphics[width=5cm,clip]{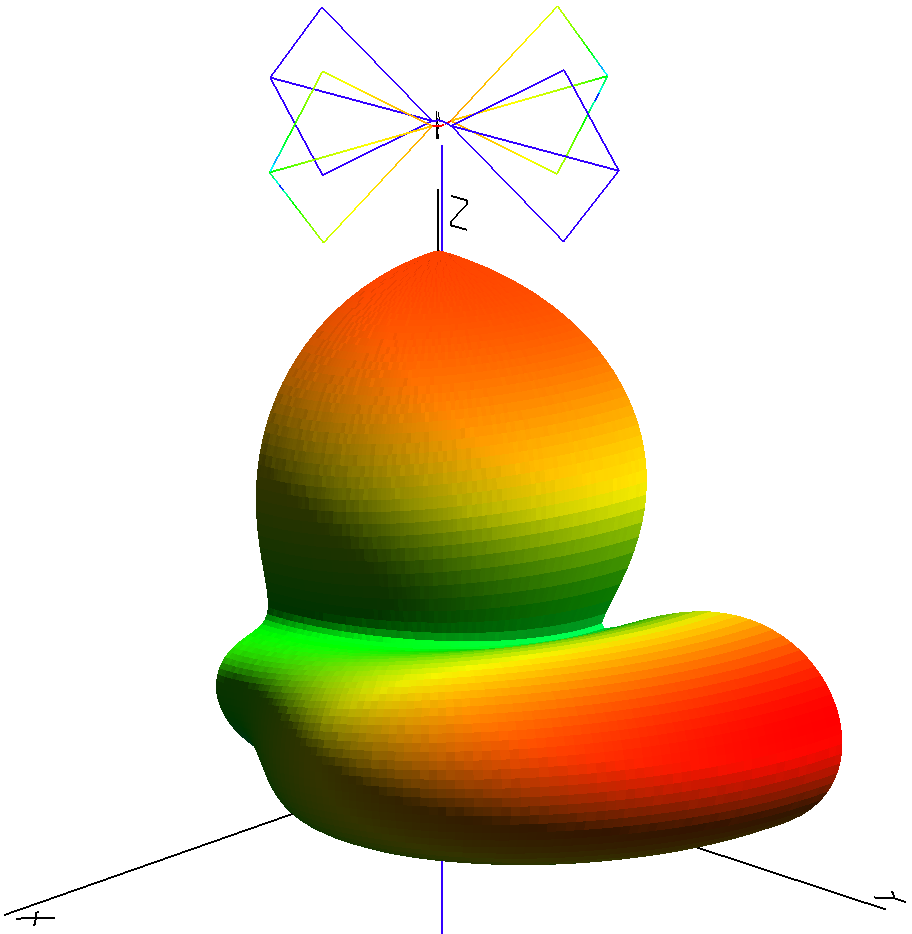}
\includegraphics[width=7cm,clip]{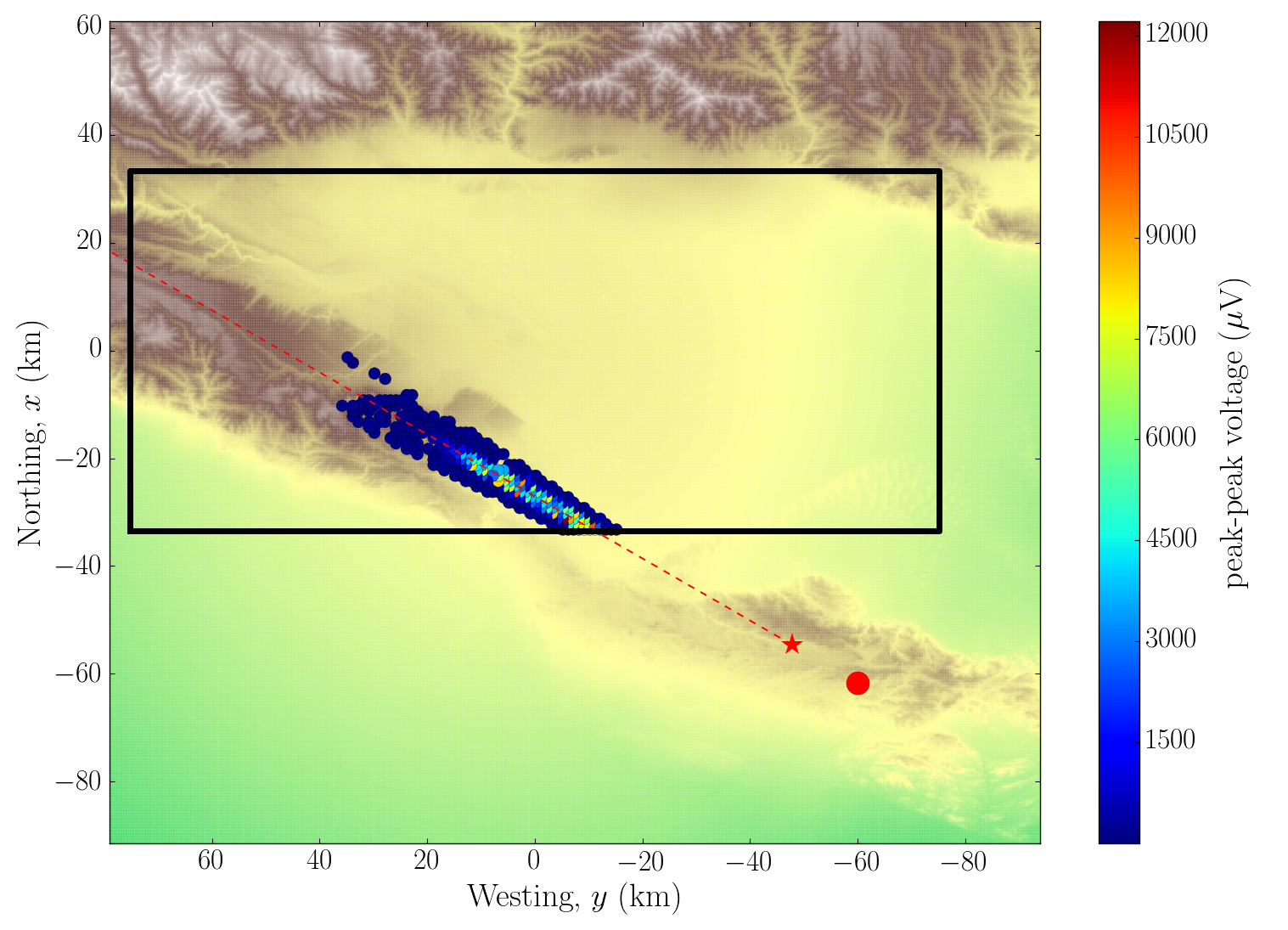}
\caption{Left: NEC4 simulation of the HorizonAntenna gain as a function of direction. Right: One simulated neutrino event displayed over the ground topography of the simulated area. The large red circle shows the position of the tau production and the red star, its decay. The
dotted line indicates the shower trajectory. Circles mark the positions of triggered antennas. The color code represents the peak-to-peak voltage amplitude of the antennas. The limits of the simulated detector are indicated with a black line. }
\label{sim}
\end{figure}

In order to estimate the potential of the GRAND detector for the detection of cosmic neutrinos, an end-to-end simulation chain was developped, composed mostly of computation-effective tools we developed to take into account the very large size of the detector and its complex topography.
\begin{itemize} 
\item The first element of the simulation chain is DANTON\,\cite{DANTON:note}, a 3-D Monte-Carlo simulation of the neutrino propagation and interactions embeded in a realistic implementation of the ground topography. A back-tracking mode is also implemented in DANTON, reducing the computation time by several orders of magnitude for neutrino energies below 10$^{18}$\,eV. 
\item The radio emission induced by each simulated tau decay is computed in our simulation chain through a semi-analytical treatment called {\it radiomorphing}. This method, detailed in \cite{Zilles:2018kwq}, allows to determine the radio signal induced by any shower  at any location through analytical operations on simulated radio signals from one single reference shower. Radiomorphing allows a gain of two orders of magnitudes in computation time compared to a standard simulation, for a relative difference of the signal amplitude below 10\% on average.
\item A specific design was proposed for the GRAND antenna. This so-called {\it HorizonAntenna} is composed of 3 arms, allowing for a complete determination of the wave polarization. Placed 5\,m above ground, with a design optimized for the 50-200\,MHz frequency range, its sensitivity to horizontal signals is excellent. The HorizonAntenna response to EAS radio signals was simulated with the NEC4 code\,\cite{NEC4} (see figure \ref{sim}) and integrated in the simulation chain. 
\item The final step of the treatment is the trigger simulation: it requires that, for at least 5 units in one 9-antennas square cell, the peak-peak amplitude of the voltage signal at the output of the antennas is larger than 30\,$\mu$V (twice the expected stationnary background noise) in an agressive scenario, or 75\,$\mu$V (five times the expected stationnary background noise) in a conservative one .
\end{itemize}
This simulation chain was run over a 10\,000\,km$^2$ area, with 10\,000 antennas deployed along a square grid of 1\,km step size in an area of the TianShan mountain range, in the XinJiang Autonomous Province (China). This setup is displayed in figure \ref{sim} together with one simulated event. The 3-year 90\% C.L. sensitivity limit derived from this simulation is presented in figure \ref{limit} and the implications on the science goals achievable by GRAND are detailed in section \ref{science}.

\subsubsection{Reconstruction performance}
\label{recons}
Reconstruction of the direction of origin, energy and nature of the primary particle from the radio data have now reached performances comparable to standard technics \cite{Buitink:2016nkf,Bezyazeekov:2015rpa,Aab:2016eeq}. A key issue for GRAND will be to achieve similar results from nearly horizontal air showers detected with radio antennas only. Demonstrating this will be one of the goal of the GRANDProto300 experiment (see section \ref{GP300}). Before that, simulation studies are used to evaluate and optimize the reconstruction performance of GRAND.

We reconstructed in particular the direction of origin of neutrino-induced air showers simulated with the ZHAireS code\,\cite{Zhaires:2012} over a GRAND-like array deployed on a toy-model topography, corresponding to a plane detector area facing the shower with a constant slope of 10$^{\circ}$ w.r.t. the horizontal. Applying a basic plane-wave hypothesis reconstruction to these data yields an average reconstruction error of a few fractions of a degree. The different antenna heights allow to achieve such resolution even for horizontal trajectories. A hyperbolic wavefront is presently being implemented, and may yield even better results, according to our understanding of air shower radio wavefront structure\,\cite{Corstanje:2014waa}. 

The method of \cite{Buitink:2014eqa} has been implemented in order to reconstruct the maximum of development of cosmic-ray induced showers on a GRAND-like array. It yields resolutions on $X_{max}$ better than 40\,g$\cdot$cm$^{-2}$ provided that the shower core position is known. 

These two preliminary results are encouraging signs that the large size of GRAND events compensate the handicaps of a sparse array and very large zenith angles. Reconstruction methods are presently being refined with a goal of 0.1$^{\circ}$ for the angular resolution and 20\,g$\cdot$cm$^{-2}$ for the $X_{max}$ resolution.

\subsection{GRAND science case}
\label{science}
\subsubsection{Ultra-high energy neutrinos}
The sources, production and nature of the particles with the highest energies in the Universe are still a mystery, despite decades-long experimental efforts. Ultra-high energy neutrinos could be an extremely valuable tool to answer this question: thanks to their very low interaction probability and neutral charge, these particles indeed travel unimpeded from their sources over cosmological distances. Besides, their production is intrisically linked to that of Ultra-High Energy Cosmic Rays (UHECRs), be it at the source itself or along the UHECRs journey through the Universe. 

\begin{figure}[h]
\centering
\includegraphics[width=6cm,clip]{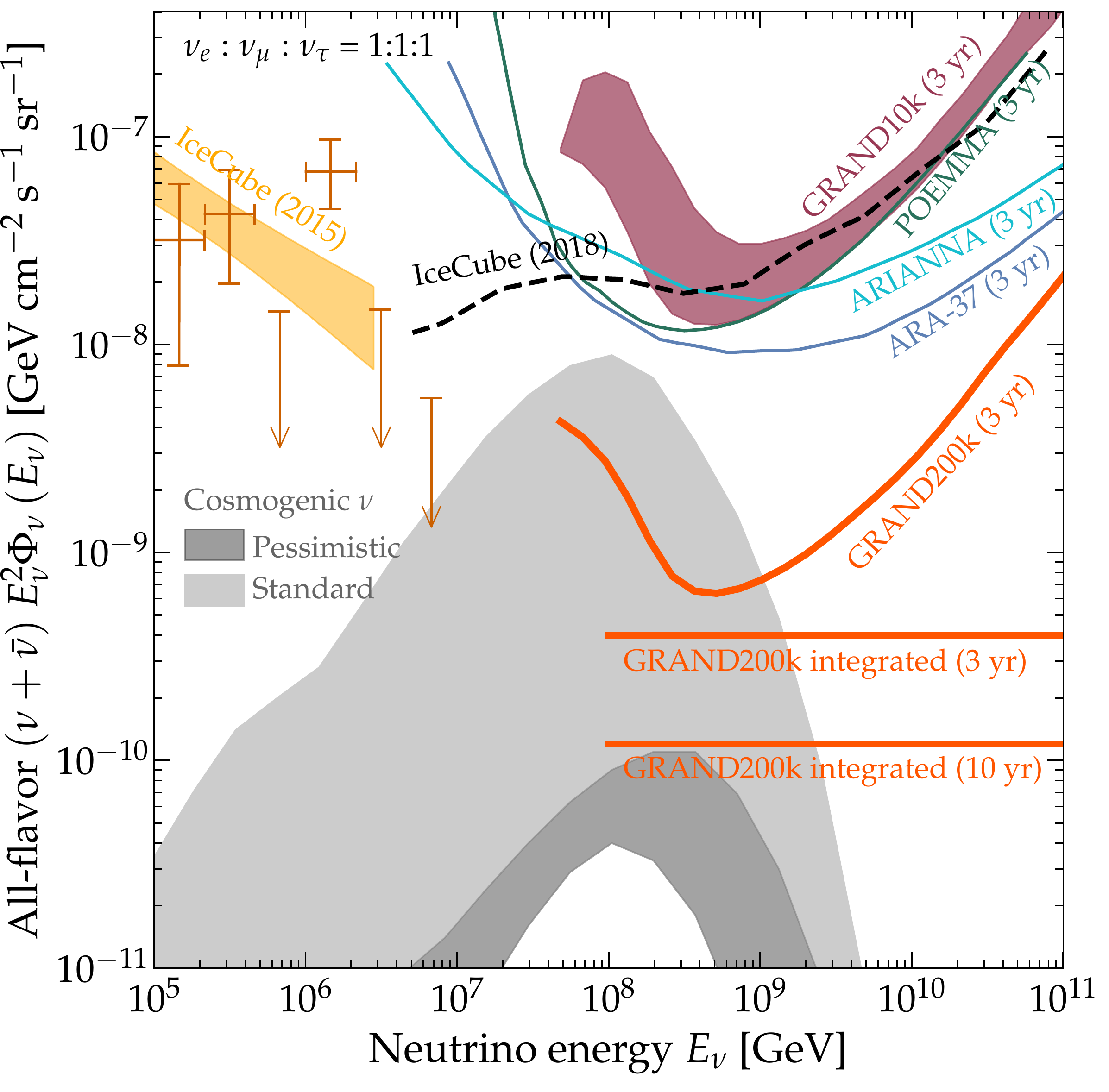}
\includegraphics[width=6cm,clip]{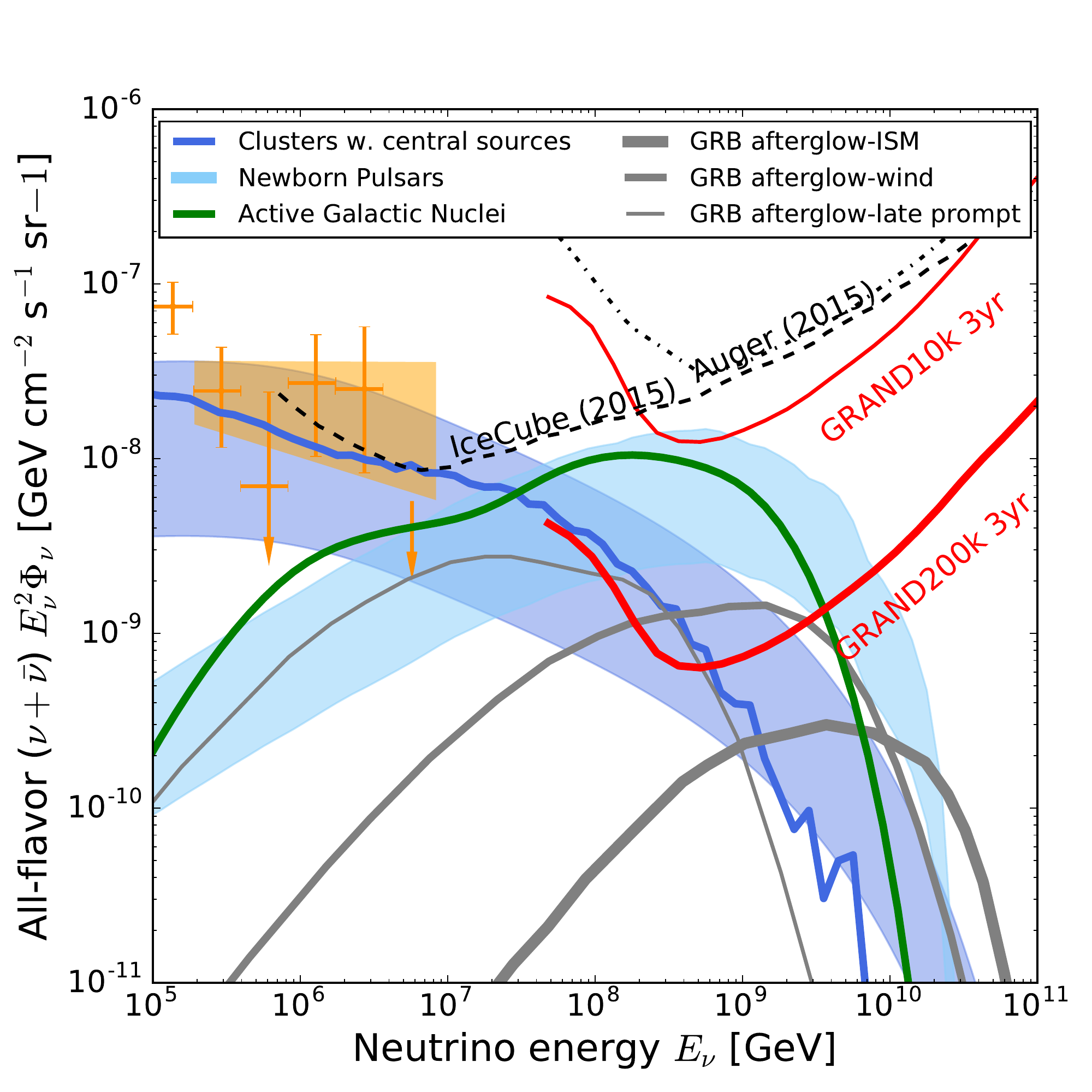}
\caption{Left: cosmogenic neutrinos flux expectations derived from the latest results of the Pierre Auger Observatory \cite{AlvesBatista:2018zui} superimposed to the differential sensitivity limit derived from the 10\,000 antennas simulation presented in section \ref{simu} ("GRAND10k", purple area) and the extrapolation for the 20-times larger GRAND array ("GRAND200k", orange line). Right: expected neutrino fluxes produced at the source for different types of sources superimposed to the GRAND10k and GRAND200k 3-years sensitivity limits. Taken from \cite{WP}.}
\label{limit}
\end{figure}

In the latter case, UHECRs interactions with the cosmic microwave background produce {\it cosmogenic} neutrinos\,\cite{GZK} which differential flux depend mostly on the astronomical evolution of the sources and the UHECRs composition and energy spectrum. A collection of 20 subarrays such as the one used in the simulation presented in section \ref{simu} would yield a neutrino sensitivity for GRAND allowing to probe the full range of expected fluxes of cosmogenic neutrinos\,\cite{AlvesBatista:2018zui} within 10 years, as illustrated in figure \ref{limit}.

For several source candidates of UHECRs, models predict fluxes of neutrinos larger than those of cosmogenic origin (see figure \ref{limit}). The GRAND sensitivity will thus allow to probe neutrino production at the source while its expected excellent angular resolution (see section \ref{recons}) will open the path for ultra-high energy neutrino astronomy.

\subsubsection{UHECRs and gamma rays}
According to preliminary simulations, GRAND will benefit from a 100\% detection efficiency for cosmic rays with zenith angles larger than 70$^{\circ}$ and energies above 10$^{18}$\,eV\,\cite{WP}. This will yield an exposure 15 times larger than the Pierre Auger Observatory. This, together with a field of view covering both Northern and Southern hemispheres, will make GRAND an excellent tool to study the end of the UHECRs spectrum. 

If a 20\,g$\cdot$cm$^{-2}$ resolution can be achieved on the $X_{max}$ measurement --- a realistic goal given present experimental results \cite{Buitink:2016nkf,Bezyazeekov:2015rpa} and preliminary simulations results, see section \ref{recons}--- GRAND will also be able to discriminate UHECRs of hadronic origin from UHE gamma rays. Non-detection of cosmogenic gamma-rays produced by photo-pion interaction of UHECRs with the CMB within 3 years of operation of GRAND would then exclude a light composition of UHECRs, while detection of UHE gamma rays from nearby sources would on the other hand probe the diffuse cosmic radio background\,\cite{Fixsen:1998kq} for instance.

\subsubsection{Fast Radio Bursts}
By incoherently adding the signals from the large number antennas in a subarray, GRAND will also be able to detect a 30 Jy fast radio burst with a flat frequency spectrum\,\cite{WP}. Moreover, as incoherent summing preserves the wide field of view of a single antenna, GRAND may be able to detect several hundreds of FRBs per days. In addition, detection of a single FRB by several subarrays would allow to reconstruct the direction of origin of the signal.

\subsection{The path to GRAND}
GRAND will perform standalone radio detection of air showers. This is a challenge, as measurements have shown that, outside polar areas, the rate of transient radio signals due to background sources (high voltage power lines or transformers, planes, thunderstorms, etc.) dominates that of EAS in the tens of MHz frequency range by several orders of magnitude, a statement that will be even truer for neutrino-induced showers. Two questions naturally arise from this observation: how to collect and identify EAS events, and how to single out neutrino-induced events among them? Below we explain how we expect to tackle these two issues.

\subsubsection{Neutrino events identification}
At energies beyond 10$^{16}$\,eV, cosmic rays are expected to induce air showers at a rate larger than neutrinos by several orders of magnitudes. Yet, selecting events with trajectories reconstructed below the horizon will allow to reject a huge fraction of them, while measuring the $X_{max}$ position (larger than 100\,km to ground for a cosmic ray with zenith angle larger than 80$^{\circ}$) will provide another very powerful discrimination tool. Reconstruction performance will therefore be of key importance to identify succesfully neutrino-induced events. 

\subsubsection{Standalone radio-detection of air showers}
\label{autonomous}

We believe that achieving a $\sim$100\% detection efficiency of EAS combined with a $\sim$100\% rejection of background is possible, taking into account the following elements:

\begin{itemize}
\item The quality of the radio environment of the detection site is of paramount importance: it is first necessary that the frequency spectrum is clean in the targeted frequency range. For the site selection of the GRANDProto300 experiment for example (see section \ref{GP300}), we request a maximum of three continuous wave emitters, and an integrated power of the spectrum not larger than twice the irreductible level due to Galactic + thermal ground emission over the targeted frequency range. More important, the rate of transient events should be evaluated with the same antenna as the one that will be used in the setup before considering deployment. In the case of GRANDProto300, a site with a trigger rate below 1\,kHz for a 6$\sigma$ threshold is considered as acceptable, where $\sigma$ is the stationnary noise level in the 50-200\,MHz range. Nine distinct sites have been evaluated for this experiment: six of them comply with the defined request in remote, mountainous areas of Western or Southern China, where major background sources are screened by mountains.

\item Even in the quietest sites, the rate of transient radio events is typically a few tens to hundreds of Hz\,\cite{Charrier:2018fle}. The DAQ system of the experiment has to be designed accordingly, in order to guarantee a 100\% live time of the acquisition system. 

\item Digital radio detection of EAS has been an active field since two decades now, but no large-scale effort of autonomous radio detection has been initiated yet: trigger algorithms remain extremely basic, very often limited to signal-over-threshold logics, and above all, self-triggered radio data is very scarce. We believe that much better can be done. To achieve this, a dedicated self-triggered radio detector has to be set up and used as a testbench to develop, test and optimize autonomous radio detection. This is one of the primary goals of the GRANDProto300 experiment, presented in the next section. 

\item Once data from all contributing antennas have been collected and written to disk, it will be possible to discriminate background event from EAS in an offline analysis based on the distinct features of the two types of signals. In particular, it has been shown already that EAS radio signatures clearly differ from background events: for example their time traces are usually much shorter\,\cite{Barwick:2016}, while their amplitude\,\cite{Nelles:2014dja} and polarization\,\cite{CODALEMA:2017} patterns at ground are very specific. Taking advantage of these specific features thus makes it possible to perform a very efficent rejection of the background signals from radio data only\,\cite{Charrier:2018fle}. Identification of EAS from radio only data is therefore not a physics issue, but a technical one.

\end{itemize}


\section{The GRANDProto300 experiment}
\label{GP300}
The GRANDProto300 (GP300) experiment will be an array of 300 detection units deployed in 2020 over 200\,km$^2$ in a $\sim$3\,000\,m-high, radio-quiet area on the borders of the Gobi desert and the Tibetan plateau. GP300 will be a pathfinder for the next stages of the GRAND project, and in particular for GRAND10k ---the first GRAND subarray of 10\,000 units--- expected to be deployed in 2025. The design of the GRAND detection unit and DAQ system will be finalized during the GRAND10k phase, and duplicated over the 20 subarrays which will compose the 200\,000-km$^2$-large GRAND detector. The deployment of GRAND is expected to be finalized in the early 2030's. 

Below we present the GP300 setup and its science goals. They are specific to the experiment, making GP300 a standalone project, meaningful even outside of the GRAND sequence.

\subsection{The GRANDProto300 setup}
The details of the GP300 detector layout will be defined once the exact location of the detector is set, in order to take maximal advantage of its topography. Its baseline design consists in three nested arrays following a square grid of 190 antennas with 1\,km step size ---the GRAND detector layout--- 500\,m step size (90 antennas) and 200\,m step size (30 antennas). 

The GP300 detection unit will be composed of the {\it HorizonAntenna} specifically designed for GRAND (see section \ref{simu}) and succesfully tested on site in summer 2018. This antenna will be placed atop a 5\,m wooden pole such as the ones used for the transportation of electrical power. The signals from the three antenna arms will be fed into 50-200\,MHz passive filters followed by 500\,MHz, 14-bits ADCs. A FPGA combined to a CPU will allow for local on-the-fly treatment of the data (filtering of continuous wave emitters, rejection of transient signals with time traces not compatible with EAS, etc). The timestamps of selected transient signals will be obtained from GPS and sent to the central DAQ over WiFi communication. For transient signals triggering five units or more, 3-$\mu$s-long time traces will then be retrieved and written to disk for offline analysis. Extrapolations of the TREND results indicate that the rate of  such events will be a few mHz only\,\cite{Martineau-Huynh:2017bpw}, but this DAQ system will allow for a 10\,Hz, a safe margin. The power consumption of the detection unit will be 10\,W. A 100\,W solar pannel will thus allow for a 24/7 acquisition time.

\subsection{Science goals of the GRANDProto300 experiment}
\label{GP300science}
GP300 will primarly serve as a demonstrator for the GRAND detection concept. In a first phase, the large amount of self-triggered data collected with GP300 will be used to refine and optimize the offline algorithms identifying EAS from their time traces shapes, amplitude and polarization patterns or frequency content (see section \ref{autonomous}). The algorithms developped with simulated data to reconstruct the primary particles informations will be applied to the selected events in order to build the sky distribution, energy spectrum and elongation rate of the candidate EAS. GP300 being too small to allow for the detection of neutrinos, these events should correspond to cosmic rays in the energy range $10^{16.5}-10^{18}$\,eV, and the comparison of these results to those obtained by other experiments will allow to quantify the performances of GP300 on the selection and reconstruction of EAS. In a second phase, EAS selection algorithms ---based on the offline treatments performed in the first phase, or on other promising technics, such as machine learning\,\cite{FuhrerARENA:2018}--- will be tested online, in order to minimize the rate of events transfered to the DAQ in the perspective of GRAND10k.

\begin{figure}[h]
\centering
\includegraphics[width=0.5\columnwidth, trim=2.5cm 7cm 3.5cm 7cm, clip=true]{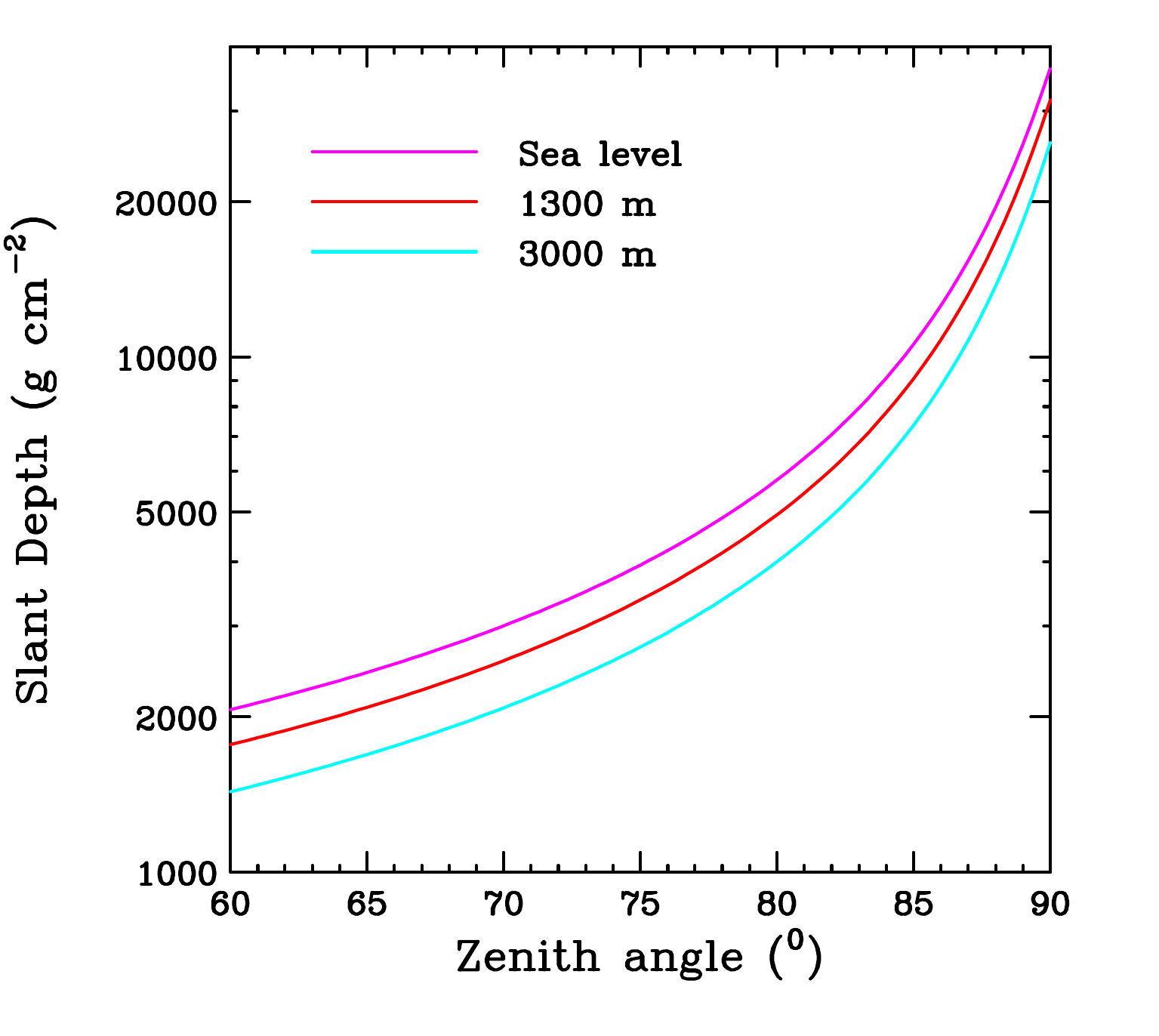}
\includegraphics[width=0.45\columnwidth, trim=0cm 0cm 1cm 1.5cm, clip=true]{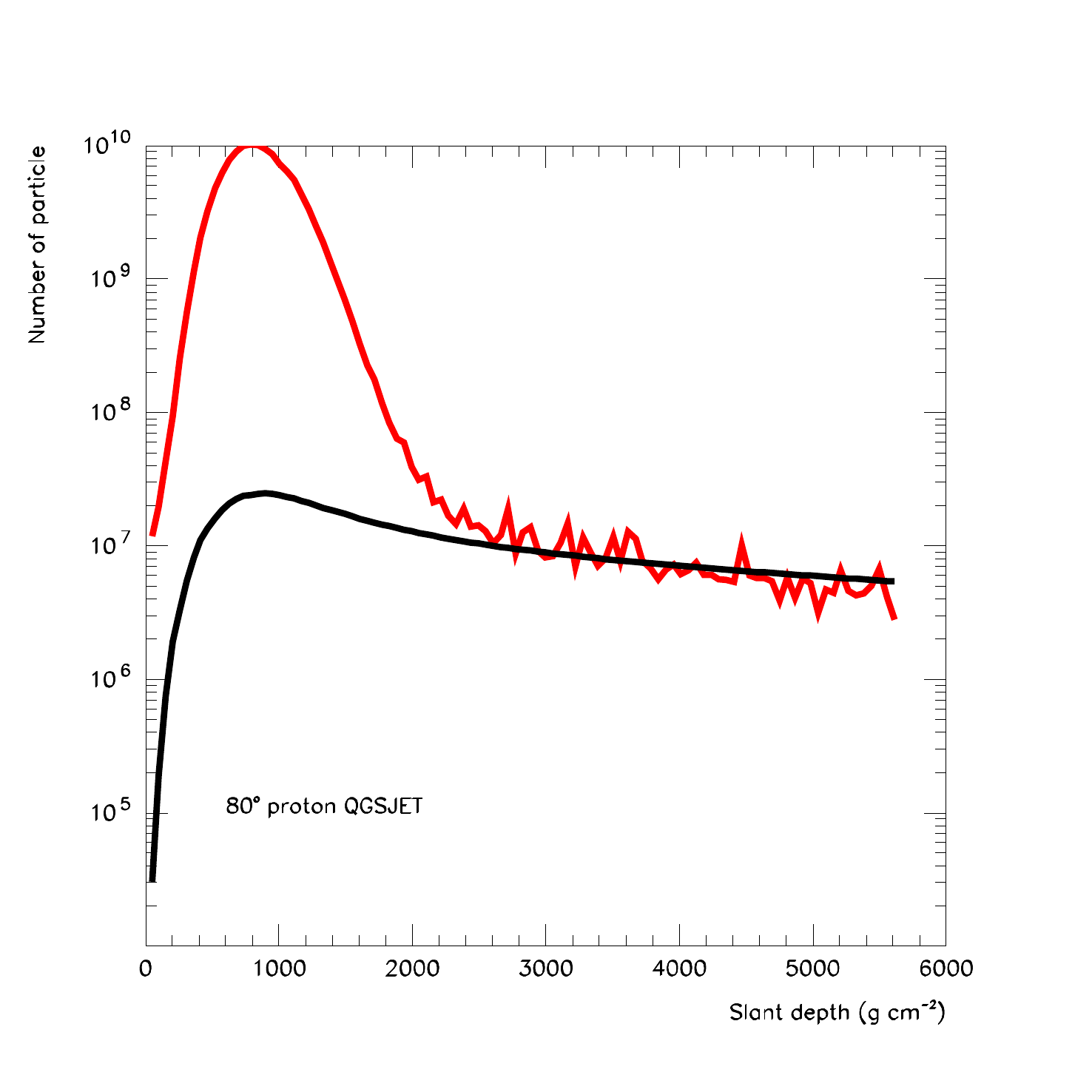}
\caption{Left: slant depth as a function of zenith angle for three different altitudes: from top to bottom sea level, 1300 and 3000~m. Right: depth development of electromagnetic (red) and muonic (black) components of a typical shower induced by a proton (qualitatively identical to that induced by a nucleus). The electromagnetic component for slant depth larger than 2000 g$\cdot$cm$^2$ is dominated by a halo produced by the decay of low energy muons, bremsstrahlung and pair production, not detectable by particle detectors. Taken from \cite{Zas:2005zz}.}
\label{muons}
\end{figure}

The GP300 radio array will be complemented by an array of particle detectors, optimized for the detection of air showers with energies above 10$^{16.5}$\,eV and zenith angles larger than 70$^{\circ}$. The electromagnetic content of these very inclined showers is fully absorbed in the atmosphere (see figure \ref{muons}), leading to a pure measurement of the muon content of the showers thanks to this particle detector array. The radio array, for its part, measures the electromagnetic part of the shower only. GP300 will therefore be the only setup performing a direct separation of the two components of the showers at the detector level itself. This is of particular interest between $10^{16.5}$ and $10^{18}$\,eV, an energy range below which a satisfying agreement between simulations and experimental data is observed\,\cite{Fomin:2016kul}, while a significant excess of muons is measured at higher energies\,\cite{Aab:2014pza} (see also \cite{dembinski} for more details on this issue). The determination of the shower energy and maximum of shower development from the radio data, combined with the measurement of the muon content by the particle detector array, on a shower-to-shower basis would provide a very interesting insight in the details of hadronic interactions at the highest energies and hopefully help us to better understand and interprete the measurements of UHECRs. GP300 will also be a suitable setup to test alternative radio-based methods to measure the nature of the primaries \cite{Billoir:2015cua}.

GP300 will also be a very effective instrument to study cosmic rays between $10^{16.5}$ and $10^{18}$\,eV: after one year of observation, it will have recorded more than 10$^5$ cosmic rays in this energy range. This large statistics, combined with the precise measurement of energy and mass composition, will allow to infer the distribution of arrival directions of light and heavy primaries separately and their variation as a function of energy \cite{WP}. This will place GP300 in a privileged position to study the transition between cosmic rays of Galactic and extragalacitc origin, expected to occur between $10^{17}$ and $10^{18}$\,eV\,\cite{Dawson:2017rsp}.

Finally GP300 will search for Giant Pulses\,\cite{Eftekhari:2016fyo}: simulations show that the incoherent summing of the signals of its 300 antennas will allow a detection of giant pulses from the Crab\,\cite{WP}, while the very wide beam formed by uncoherent summing will allow for full sky survey of sources of similar intensities. GP300 will also study the Epoch of Reionization: an absolute calibration of 30 antennas at the 1\,mK level ---a goal reachable\,\cite{WP}--- and a summation of their signals would allow to measure the temperature of the sky with a precision large enough to identify the absorption feature due to the reionization of hydrogen by the first stars, expected below 100\,MHz.

\section{Conclusion}
A detailed, robust and reliable simulation chain demonstrates that the 200\,000\,km$^2$ GRAND detector will reach a sensitivity  allowing the detection of neutrinos of cosmogenic origin or directly emitted by the source. This, combined to its exposure to UHECRs and UHE gamma rays, will make GRAND a very powerfull tool to study the origin of UHECRs. The GRANDProto300 experiment, a setup of 300 antennas covering a 200\,km$^2$ area,  will be deployed in 2020 in order to demonstrate that the autonomous detection of very inclined showers is possible, thus validating GRAND detection principle. GRANDProto300 will not only be a pathfinder for GRAND: it will also be a self-standing experiment thanks to its rich and appealing science case. 

\bibliography{biblio}

\begin{thebibliography}{33}

\bibitem{Huege:2016veh}
T.~Huege, Phys.\ Rept. \textbf{620}, 1 (2016), \texttt{1601.07426}

\bibitem{Schroder:2016hrv}
F.G. Schr{\"o}der, Prog. Part. Nucl. Phys. \textbf{93}, 1 (2017),
  \texttt{1607.08781}

\bibitem{WP}
J.~Alvarez-Mu\~niz et~al. (GRAND) (2018), \texttt{1810.09994}

\bibitem{Kahn206}
F.~Kahn, I.~Lerche, Proceedings of the Royal Society of London A: Mathematical,
  Physical and Engineering Sciences \textbf{289}, 206 (1966),
  \texttt{http://rspa.royalsocietypublishing.org/content/289/1417/206.full.pdf}

\bibitem{Allan:1971}
H.~Allan, Progress in elementary particle and cosmic ray physics \textbf{10}
  (1971)

\bibitem{Ardouin:2005qe}
D.~Ardouin et~al., Nucl. Instrum. Meth. \textbf{A555}, 148 (2005),
  \texttt{astro-ph/0504297}

\bibitem{Falcke:2005tc}
H.~Falcke et~al. (LOPES), Nature \textbf{435}, 313 (2005),
  \texttt{astro-ph/0505383}

\bibitem{Fargion:2000iz}
D.~Fargion, Astrophys.\ J. \textbf{570}, 909 (2002), \texttt{astro-ph/0002453}

\bibitem{DANTON:note}
V.~Niess, O.~Martineau-Huynh (2018), \texttt{1810.01978}

\bibitem{Zilles:2018kwq}
A.~Zilles, O.~Martineau-Huynh, K.~Kotera, M.~Tueros, K.~de~Vries,
  W.~Carvalho~Jr., V.~Niess, N.~Renault-Tinacci, V.~Decoene (2018),
  \texttt{1811.01750}

\bibitem{NEC4}
G.~Burke (1992), {\tt
  http://physics.princeton.edu/$\sim$mcdonald/examples/\\NEC\_Manuals/NEC4UsersMan.pdf}

\bibitem{Buitink:2016nkf}
S.~Buitink et~al., Nature \textbf{531}, 70 (2016), \texttt{1603.01594}

\bibitem{Bezyazeekov:2015rpa}
P.A. Bezyazeekov et~al., Nucl.\ Instrum.\ Meth.\ A \textbf{802}, 89 (2015),
  \texttt{1509.08624}

\bibitem{Aab:2016eeq}
A.~Aab et~al. (Pierre Auger), Phys. Rev. Lett. \textbf{116}, 241101 (2016),
  \texttt{1605.02564}

\bibitem{Zhaires:2012}
J.~{Alvarez-Mu{\~n}iz}, W.R. {Carvalho}, E.~{Zas}, Astroparticle Physics
  \textbf{35}, 325 (2012), \texttt{1107.1189}

\bibitem{Corstanje:2014waa}
A.~Corstanje et~al., Astropart. Phys. \textbf{61}, 22 (2015),
  \texttt{1404.3907}

\bibitem{Buitink:2014eqa}
S.~Buitink et~al., Phys.\ Rev.\ D \textbf{90}, 082003 (2014),
  \texttt{1408.7001}

\bibitem{GZK}
G.T. Zatsepin, V.A. Kuzmin, JETP Lett. \textbf{4}, 78 (1966), [Pisma Zh.\
  Eksp.\ Teor.\ Fiz.\ 4, 114(1966)]

\bibitem{AlvesBatista:2018zui}
R.~Alves~Batista, R.M. de~Almeida, B.~Lago, K.~Kotera (2018),
  \texttt{1806.10879}

\bibitem{Fixsen:1998kq}
D.J. Fixsen, E.~Dwek, J.C. Mather, C.L. Bennett, R.A. Shafer, Astrophys. J.
  \textbf{508}, 123 (1998), \texttt{astro-ph/9803021}

\bibitem{Charrier:2018fle}
D.~Charrier et~al. (2018), \texttt{1810.03070}

\bibitem{Barwick:2016}
S.W. Barwick et~al., Astroparticle Physics \textbf{90}, 50 (2017),
  \texttt{1612.04473}

\bibitem{Nelles:2014dja}
A.~Nelles et~al., Astropart. Phys. \textbf{65}, 11 (2015), \texttt{1411.6865}

\bibitem{CODALEMA:2017}
D.~Garc\'ia-Fern\'andez, ed., \emph{{The CODALEMA/EXTASIS experiment:
  Contributions to the 35th International Cosmic Ray Conference (ICRC 2017)}}
  (2017), \texttt{1710.02487}

\bibitem{Martineau-Huynh:2017bpw}
O.~Martineau-Huynh et~al. (GRAND), EPJ Web Conf. \textbf{135}, 02001 (2017),
  \texttt{1702.01395}

\bibitem{FuhrerARENA:2018}
F.~F{\"u}hrer, T.~Charnock, A.~Zilles, M.~Tueros, \emph{{Towards online
  triggering for the radio detection of air showers using deep neural
  networks}} (2018), \texttt{1809.01934}

\bibitem{Zas:2005zz}
E.~Zas, New J.\ Phys. \textbf{7}, 130 (2005), \texttt{astro-ph/0504610}

\bibitem{Fomin:2016kul}
{\relax Yu}.A. Fomin, N.N. Kalmykov, I.S. Karpikov, G.V. Kulikov, M.{\relax
  Yu}. Kuznetsov, G.I. Rubtsov, V.P. Sulakov, S.V. Troitsky, Astropart. Phys.
  \textbf{92}, 1 (2017), \texttt{1609.05764}

\bibitem{Aab:2014pza}
A.~Aab et~al. (Pierre Auger), Phys. Rev. \textbf{D91}, 032003 (2015), [Erratum:
  Phys. Rev.D91,no.5,059901(2015)], \texttt{1408.1421}

\bibitem{dembinski}
H.~Dembinski, \emph{these proceedings} (2018)

\bibitem{Billoir:2015cua}
P.~Billoir, M.~Settimo, M.~Blanco, Astropart.\ Phys. \textbf{74}, 14 (2016),
  \texttt{1508.04354}

\bibitem{Dawson:2017rsp}
B.R. Dawson, M.~Fukushima, P.~Sokolsky (2017), \texttt{1703.07897}

\bibitem{Eftekhari:2016fyo}
T.~Eftekhari, K.~Stovall, J.~Dowell, F.K. Schinzel, G.B. Taylor, Astrophys.\ J.
  \textbf{829}, 62 (2016), \texttt{1607.08612}

\end{thebibliography}
%
%
%
%

\end{document}